# Electronic Structure of ZnO nanowire

Xiao Shen

*Department of Physics and Astronomy, Stony Brook University, Stony Brook, NY 11794-3800*

Mark R. Pederson

*Complex Systems Theory Branch, Naval Research Laboratory, Washington, D.C. 20375-5000*

Jin-Cheng Zheng

*Center for Functional Nanomaterials, Brookhaven National Laboratory, Upton, NY 11973-5000*

James W. Davenport

*Computational Science Center, Brookhaven National Laboratory, Upton, NY 11973-5000*

James T. Muckerman

*Chemistry Department, Brookhaven National Laboratory, Upton, NY 11973-5000*

Philip B. Allen

*Department of Physics and Astronomy, Stony Brook University, Stony Brook, NY 11794-3800[1] and*
*Center for Functional Nanomaterials, Brookhaven National Laboratory, Upton, NY 11973-5000*

(Dated: September 29, 2006)

This paper presents two first-principles calculations on the lattice- and electronic-structures of a small-diameter infinite and truncated ZnO nanowire. The two calculations show excellent agreement with each other. For the small diameter nanowire, the lattice and electronic properties are determined by the relaxed surface. We find a similarity between the nanowire surface and the surface of the bulk material, and discuss the assignment of the rotational quantum number $m$ to the bands.

PACS number(s): 73.22.-f, 62.25.+g, 61.46.Hk

## I. INTRODUCTION

ZnO nanowires have drawn attention because of their potential for application and ease of fabrication.[1-3] Here we report first principles calculations for the geometry and electronic structure of a small diameter ZnO nanowire. The calculations were performed on an infinite wire and a truncated piece. Our results provide the small diameter limit of ZnO nanowire properties.

## II. METHOD OF COMPUTATION

### A. Infinitely long wire

We consider a small diameter nanowire formed from the wurtzite phase of ZnO. First we treat a wire which is periodic in the $z$ direction with lattice parameter $c$, containing 13 Zn and 13 O atoms in one unit cell. The unrelaxed structure of the nanowire is shown in Fig. 1(a) and (b), where Fig. 1(b) shows 4 unit cells. We assume that the surface atoms are not saturated by foreign atoms. The calculation uses the plane-wave pseudopotential method, density functional theory (DFT), and the Quantum Espresso/PWSCF code.[4] We use Vanderbilt ultrasoft pseudopotentials[5] with the Perdew-Burke-Ernzerhof (PBE) version of the generalized gradient approximations (GGA) exchange-correlation functional[6]. The Zn 4$s$ and 3$d$ states and O 2$p$ and 2$s$ states are treated as valence electrons. The cutoff of kinetic energy is set at 30 Ry, while the cutoff of charge density is set at 300 Ry. The Broyden-Fletcher-Goldfarb-Shanno (BFGS) method is used for the geometry optimizations. A 5x5x5 $k$-point mesh is used for the bulk calculation and a 1x1x10 mesh is used for the nanowire supercell calculation. During relaxation of the nanowire, a Gaussian broadening[7] with smearing parameter of 0.002Ry is used for Brillouin-zone integrations to prevent possible convergence problems due to fractionally occupied surface states. The force on each atom is converged at 0.003eV/Å for all optimizations. We arrange the ZnO nanowires in a 2D hexagonal supercell and separate the facets of adjacent nanowires by at least 7.5 Å. For such a small diameter wire, the spacing $c$ is not the same as bulk. In order to find $c$ for the optimized nanowire, we fixed $c$ at a set of different values, and relaxed the nanowire until the force on each atom is converged. Then we plot the total energy after relaxation versus $c$. A parabolic fit gives the lowest energy and the corresponding interpolated $c$ value. This method also gives the Young's modulus of the nanowire. We then fixed $c$ at that value and let the wire relax to its final configuration.

---

[1] Permanent address

### B. Truncated Wire

We also calculated a finite $(ZnO)_{52}$ nanowire, which contains 4 unit cells of the infinitely long wire as shown in Fig. 1. All-electron calculations on this truncated nanowire were performed using the NRLMOL suite of DFT codes. At each O site, a total of 7 $s$-type functions, 4 $p$-type functions and 2 $d$-type functions (39 contracted functions per atom) were used. The contracted functions were constructed from 13 primitive Gaussians with decay constants ranging from $6.12 \times 10^6$ to 0.1049. Similarly at each Zn site, a total of 10 $s$-type, 5 $p$-type and 3 $d$-type contracted orbitals were constructed from 20 optimized primitive Gaussians with decay parameters ranging between $5.008 \times 10^8$ and 0.0554. The decay parameters were optimized to minimize the PBE-GGA energy of the isolated atoms using the method discussed in Ref. 8. The nanowire was optimized using the PBE-GGA energy functional. The relaxation was done within the constraint of $C_{3V}$ symmetry.

### III. RESULTS AND DISCUSSION
#### A. Infinitely long wire
*1.Bulk Results*

Computed parameters of bulk wurtzite phase ZnO are shown in Table I. They provide a starting point for the nanowire calculation, and serve as a test of the plane-wave pseudopotential implementation. The cohesive energy is obtained by subtraction of the total energy per ZnO formula unit at the equilibrium position, from the sum of the energy of isolated Zinc and Oxygen atoms, with spin polarization, but not including the zero-point motion. The isolated atom energies are calculated by arranging them on a simple cubic supercell with lattice parameter 11 Å. The bulk modulus is obtained by fitting the total energy as a function of volume to the

Murnaghan equation of state.[9] Our results agree well with other GGA calculations and experiments.

**TABLE I.** Comparison between our bulk results and other GGA calculation and experiments. The spacing between closest Zn and O layers is $(1/2-u)c$.

|  | Present Work | Other GGA | Experiment |
|---|---|---|---|
| $a$ (Å) | 3.292 | 3.283[a] | 3.250[c] |
| $c$ (Å) | 5.309 | 5.289[a] | 5.207[c] |
| $u$ | 0.3793 | 0.378[a], 0.3790[b] | 0.3817[c] |
| $E_{coh}$(eV/f.u.) | 7.27 | 7.692[b] | 7.56[d] |
| $B_0$ (GPa) | 127 | 149[a], 133.7[b] | 142.6[e] |
| Band Gap (eV) | 0.71 | 0.75[a] | 3.4[f] |

[a] Reference 10.　　[b] Reference 11.
[c] Reference 12.　　[d] Reference 13.
[e] Reference 14.　　[f] Reference 15.

*2. Optimized Structure and Cohesive Energy*

As shown in Fig. 2, the optimized nanowire has $c$ =5.425 Å, which is 2.2% larger than the bulk value. The relaxed structure of the ZnO nanowire is shown in Fig. 1(c) and (d). The change of $c$ occurs because the atoms on the surface contract inwards in the x-y plane, reducing the in-plane distance between the center atom and outermost atom (green line in Fig. 1(c)) from 3.801 Å to 3.743 Å, thus shrinking the diameter of nanowire by 1.5%. The in-plane contraction increases the repulsion in the z direction. The in-plane contraction of surface atoms can be understood from Fig. 3, which shows the charge distribution in bulk ZnO(a), at the surface of an unrelaxed nanowire(b), and at the surface of a relaxed nanowire(c). In bulk ZnO, atoms are at equilibrium positions, the electrons are attracted by the atoms on the left. The attraction from left side is missing for a surface atom of the unrelaxed nanowire. Compared with the bulk, the charge moves inwards, and the ions will also move inwards.

The relaxed structure of the ZnO nanowire is very similar to the $(10\bar{1}0)$ surface of wurtzite ZnO[16-19]. In our optimized nanowire, the surface ZnO bonds parallel to the z axis, at position A and B in Fig. 1(c), are tilted by 8.5° and 9.6°, close to the 10.1° value in Ref. 18. Those bonds are also contracted by 6.6% and 5.8% in our calculation, close to the 7% value in Ref. 18. This similarity is easy to understand. Taking the outer-most layer of the $(10\bar{1}0)$ surface and wrapping it around z axis to form a triangle (shown as blue lines in Fig. 1(c)), gives the surface of our nanowire.

For an unrelaxed wire, in which atoms are placed as in the bulk, the cohesive energy is 6.46eV/f.u.. The cohesive energy for a relaxed wire, is 6.67eV/f.u., while for bulk the value is 7.27 eV/f.u from our calculation. Thus the formation energies $E_{wire}$-$E_{bulk}$ of unrelaxed and relaxed wire are 0.81eV/f.u. and 0.60eV/f.u., respectively. This translates into 0.88 and 0.65eV per surface Zn-O dimer, close to the 0.80 and

0.43 eV per surface Zn-O dimer given in Ref. 18.

### 3. Young's Modulus

The parabolic fit of total energy vs. $c$ gives the Young's modulus $E_3$=257 GPa of this nanowire, significantly larger than the bulk value. Recently, Chen *et al.* reported an experimental measurement of the size-dependent $E_3$ of ZnO nanowires.[20] They found that, when the diameter of the nanowire is smaller than 120nm, the Young's modulus increases significantly with decreasing diameter. Our result agrees with the experimental trend. 257 GPa can be regarded an estimated upper limit of the Young's modulus for small diameter ZnO nanowires.

### 4. Electronic Structure

The band structure of a relaxed wire is shown in Fig. 4, where the bands are grouped by rotational quantum number $m$ (see below). The total density of state (DOS) of relaxed and unrelaxed wires are shown in Fig. 5. Fig. 6 shows the main components of valence and conduction bands of the relaxed nanowire. Fig. 5 shows that the band gap increases as the wire relaxes. This relaxation may be due to either the change of the $c$ parameter or to surface reconstruction. To determine which effect accounts for the gap increase, we compared the band gap for relaxed nanowires with different fixed $c$ values. The results are shown in Table II. The band gap is not sensitive to the $c$ value, and therefore is mainly determined by the relaxation of surface atoms.

**TABLE II.** Band gap vs. $c$

| $c$ (Å) | Band gap (eV) |
|---|---|
| 5.309(unrelaxed) | 1.03 |
| 5.309 | 1.62 |
| 5.380 | 1.65 |
| 5.425 | 1.64 |
| 5.480 | 1.62 |

Our nanowire band gap, 1.64 eV, is larger than the DFT bulk value 0.71 eV. A similar result was reported by Schröer *et al.* for the surface states of ZnO $(10\bar{1}0)$ surface.[17] Those authors suggested that the unoccupied states are shifted upwards after the relaxation and the occupied states are shifted downwards. However, we can not compare our band gap with Ref. 17 numerically because in their case, the surface states and bulk states are mixed near the Gamma point. In our case, the core is small and the surface determines the electronic structure of the nanowire. A similar conclusion was drawn in Ref. 21 for the small diameter InP nanowire.

Unlike the bulk $(10\bar{1}0)$ surface, our nanowire has a 3-fold rotational symmetry. In 1D nanowires, the rotations and translations along $z$-axis commute with each other, so they can be simultaneously diagonalized.[22] The Hamiltonian is invariant under rotations by *2π/3*. The eigenvalues of the rotation operator are *exp(i2pm/3)*, with *m=-1,0,1*. The *m=-1 and m=+1* states are degenerate. Fig 3 shows bands classified according to *m* value. A few typical examples are shown in Fig. 7, where the wave functions of the bands (all at the Γ point) are plotted. The lowest unoccupied (LU) band is non-degenerate, and corresponds to *m*=0, as shown in Fig. 7(a). The next lowest unoccupied states, Fig. 7(b)(c), are degenerate, each with one node, corresponding to linear combinations of *m*=+1 and -1 states. The highest occupied (HO) state, Fig. 7(d), belongs to *m*=0, with 3 nodes coming from 3 reflection symmetries. Similar behavior exists in Ref. 21 for a triangular shape InP nanowire.

### B. Truncated Wire
#### 1. Cohesive Energy and Structure parameters.

The final geometry was 6.62 eV per formula unit lower in energy than the separated spin-polarized spherical atoms. The interior ZnO bondlength is 2.016 Å. Similarly, the interior "*c*" lengths are in the range of 5.303 (O-O) to 5.304 (Zn-Zn) Å. Based on previous experience and what is known about classical pair potentials (e.g. Lennard-Jones etc) these "*c*" lengths should be a lower bound for an infinite-length nanowire. The value 5.425 Å obtained in Section A is consistent with this. For the Zn terminated (top) end, the ZnO bonds that are parallel to the axis of symmetry are ≈ 1.969-2.188 Å for the interior and exterior axial bondlengths. Inequivalent tangential ZnO surface bonds are found to be 1.907, 1.969, 1.944 and 1.975 Å. On the oxygen terminated bottom surface, the bonds are decreased further, with inequivalent tangential surface bonds of 1.875, 1.875, 1.979 and 1.849 Å. The bonds parallel to the axis are found to be 2.223 Å for both types. In the interior of the truncated nanowire, the relaxed geometry agrees well with the infinite nanowire calculation.

#### 2. Electric dipole moment

The relaxed nanowire is 19.13 Å long and has a net dipole moment of 17.23 a.u.. This is significantly smaller than the dipole moment expected based upon a tube of +/- 2 point charges (166.8 a.u.). The reduction of dipole moment is due to incomplete charge transfer on the surface layers (discussed below).

#### 3. Electronic Structure

Fig. 8 shows the total density of states, and projections onto eight layers. They are obtained by the projecting the occupied density of states onto non-overlapping spheres using the same Gaussian-broadening as Fig. 5 and Fig. 6, with the Fermi level shifted to zero. The top surface has an outer layer of Zn atoms above a layer of O atoms. Similarly the bottom surface is defined by an outer layer of O atoms with Zn atoms above. The total density of states shows Zn (3*d*) weight in the energy range between -9.0 to -5.0 eV and O (3*p*) DOS in the energy range -5 eV to 0 eV. The projected DOS shows a similar behavior with the Zn 3*d* portions below the respective O portions. Due to the positively charged top layer and the negatively charged bottom layer, the local density of states shows a Coulomb shift in the projected DOS. As the length of the molecular crystal increases, this Coulomb shift should disappear at distance Δ*z* larger than radius away from the top and bottom surface, and is completely absent from the infinite nanowire calculation. In addition, top and bottom surfaces show spectral weight above the interior Fermi level. This causes the incomplete charge transfer which has decreased the dipole moment. The bottom surface, with open-shell oxygen atoms, shows the most prominent surfaces states. In contrast to the DOS of the infinite ZnO nanowire, the truncated nanowire displays a slightly enhanced gap (3.0 eV), with some surface states appearing in between the valence and conduction band.

Due to the internal electric fields created by the dipole moment of the truncated wire, the valence band width Zn(3*d*)-O(2*p*) is broadened. For this reason the total width agrees well, but slightly overestimates, the infinite wire calculation For example the Zn(3*d*) width is approximately 3.0 eV (last panel of Fig. 8) rather than the ~2.5 eV Zn(3*d*) width observed in the infinite wire calculations (Fig. 6). The larger bandwidth associated with the truncated wire is in stark contrast to nonpolar materials where one always expects the bandwidth to increase monotonically with the size of a crystal fragment. Indeed, it is the width of the interior Zn(3*d*) DOS (~1.5 eV) which should eventually converge to the infinite wire width ~2.5 eV). Since all of the atoms in these wires, as well as slightly wider wires, are spectroscopically observable, observation of the valence band width during and after the synthesis process provides a useful means for determining the length of a given wire. As a wire grows longer, one expects the Zn(3*d*) DOS to increase beyond that of the infinite length tube, and then eventually decrease as the effect of the tip states become neglible. Another point that might be useful for real-time monitoring of wire growth, is that the oxygen 1s-core levels show the same Coulomb shifts as the valence (Zn(3*d*)O(2*p*)) levels. Total peak shifts between the top and bottom surfaces observed in Fig. 8 is approximately 2.7 eV which is very similar to the splittings obtained in the O(1*s*) core levels of 2.9 eV. In contrast, the splitting of the Zn(1*s*) states remains small (0.46 eV).

To make a more direct comparison to the results of Sec A, Fig. 9 shows the DOS of the interior of the truncated nanowire.

### 4. Charge States of Truncated Wire

For the lowest-energy geometry, several negative charge states of the molecule have been calculated. The lowest-energy charge state is the trianion. The trianion has an energy that is 6.2 eV lower than that of the neutral truncated wire. The relatively high anionic charge is somewhat rare in isolated systems but not overly surprising since the wire is rather long in comparison to most "finite molecules".

### IV. CONCLUSION

For a small diameter ZnO nanowire, the surface atoms relax and the length of the nanowire increases. The change of length does not have significant effect on the band gap of the nanowire. Both the surface geometry and the electronic structure show the features of a bulk ($10\bar{1}0$) surface. However, unlike the bulk surface, the band structure of nanowire has an additional degeneracy related to the rotational symmetry of the nanowire and can be grouped by rotational quantum number *m*.

### ACKNOWLEDGEMET

We thank J. Mintmire and T. Sun for helpful discussions. Research at Brookhaven National Laboratory was supported by U.S. DOE under contract No. DE-AC02-98CH10886. Work at Stony Brook was supported in part by NSF grant no. NIRT-0304122, and in part by a BNL-Stony Brook Seed Grant.

---

[1] Z. Fan and J.G. Lu, J. Nanosci. Nanotech. **5**, 1561 (2005)

[2] Y. Y. Wu and P. D. Yang, J. Am. Chem. Soc. **123**, 3165 (2001)

[3] Z. L. Wang, J. Phys.: Condens. Matter **16**, R829 (2004)

[4] S. Baroni, A. Dal Corso, S. de Gironcoli, and P. Giannozzi, Plane-Wave Self-Consistent Field, http://www.pwscf.org.

[5] D. Vanderbilt, Phys. Rev. B **41**, R7892 (1990).

[6] J. P. Perdew, K. Burke, and M. Ernzerhof, Phys. Rev. Lett. **77**, 3865 (1996)

[7] C. L. Fu and K. M. Ho, Phys. Rev. B **28**, 5480 (1983)

[8] D. Porezag and M.R. Pederson Phys. Rev A **60** 2840 (1999).

[9] F. D. Murnaghan, Proc. Natl. Acad. Sci. USA **30**, 244 (1944)

[10] P. Erhart, K. Albe, and A. Klein, Phys. Rev. B **73**, 205203 (2006).

[11] J. E. Jaffe, J. A. Snyder, Z. Lin, and A. C. Hess, Phys. Rev. B **62**, 1660 (2000).

[12] E. H. Kisi and M. M. Elcombe, Acta Cryst. **C45**, 1867 (1989).

[13] *CRC Handbook of Chemistry and Physics*, 58th ed. (CRC, Boca Raton, 1977)

[14] S. Desgreniers, Phys. Rev. B **58**, 14102 (1998)

[15] *Numerical Data and Functional Relationships in Science and Technology*, Landolt-Börnstein, New Series Group III, Vol. 17b, edited by K.-H. Hellwege and O. Madelung (Springer, New York, 1982).

[16] U. Diebold, L. V. Koplitz and O. Dulub, Appl. Surf. Sci. **237**, 336 (2004)

[17] P. Schröer, P. Krüger, and J. Pollmann, Phys. Rev. B **49**, 17092 (1994)

[18] B. Meyer and D. Marx, Phys. Rev. B **67**, 035403 (2003)

[19] A. Wander and N. M. Harrison, Surf. Sci. **457**, L342 (2000).

[20] C. Q. Chen, Y. Shi, Y. S. Zhang, J. Zhu and Y. J. Yan, Phys. Rev. Lett. **96**, 075505 (2006)

[21] T. Akiyama, K. Nakamura, and T. Ito, Phys. Rev. B **73**, 235308 (2006)

[22] E. Chang, G. Bussi, A. Ruini, and E. Molinari, Phys. Rev. Lett. **92**, 196401 (2004)

[23] The ball-stick graphs, charged density and wave function plots are generated by XCrysden graphical package. A. Kokalj, J. Mol. Graphics Modeling **17** (1999) 176–179. Code available http://www.xcrysden.org/

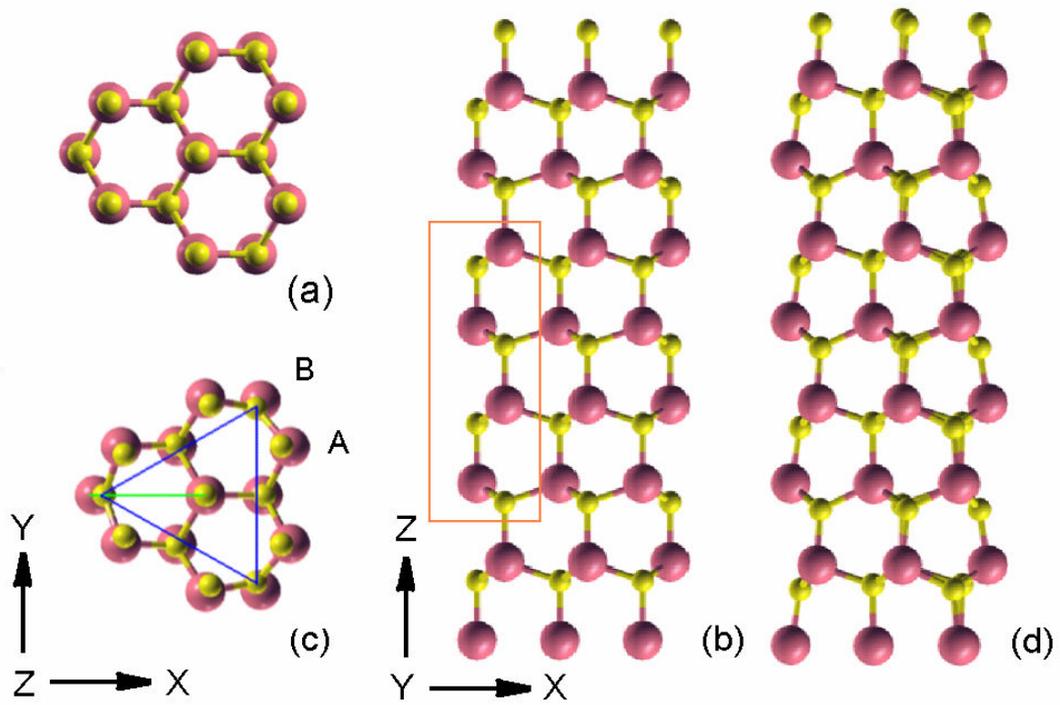

**FIG. 1.** Unrelaxed and relaxed structure of ZnO nanowire. The axis of the nanowire is in z-direction. Yellow balls represent zinc atoms and red balls represent oxygen atoms. 13 Zn atoms from 2 layers are visible in (a) and (c).

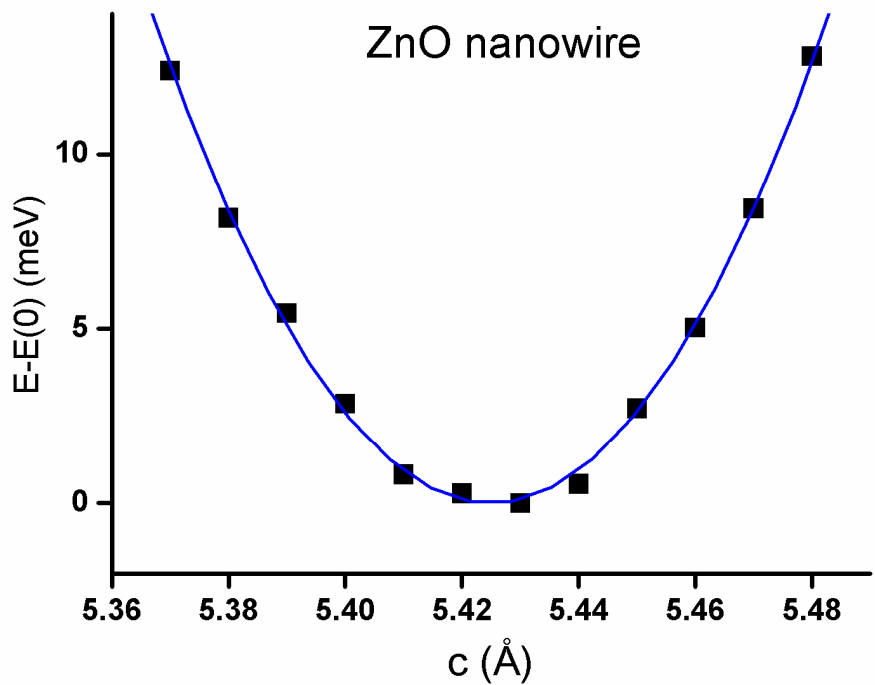

**FIG. 2.** Total energy of ZnO nanowire VS. fixed lattice parameter *c*

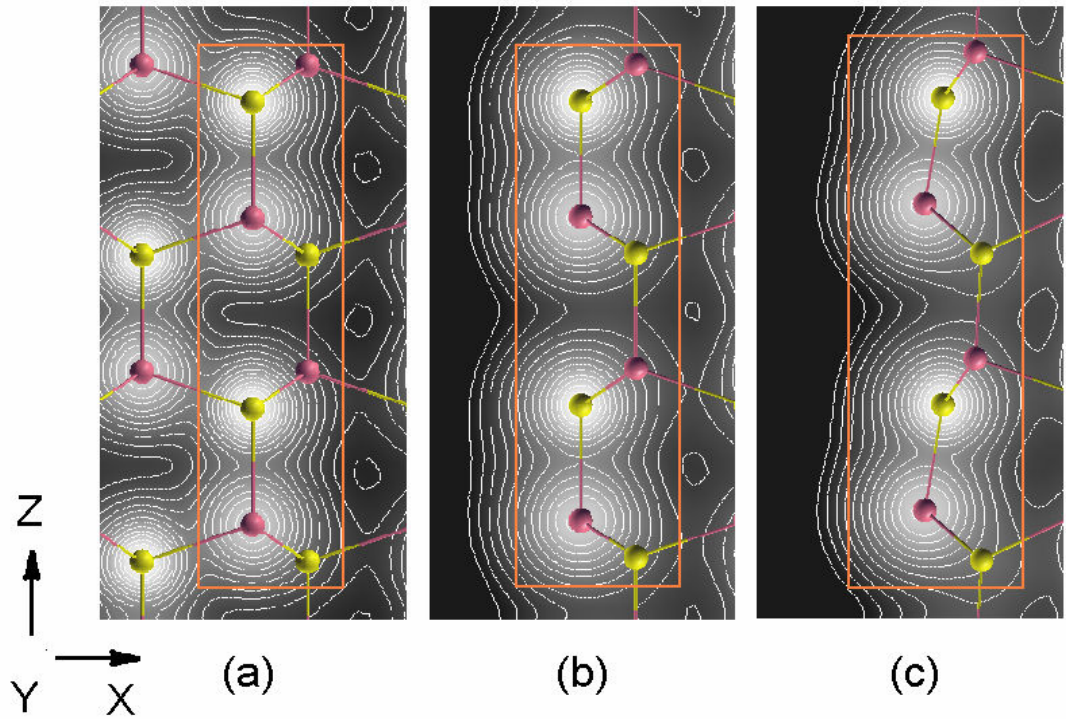

**FIG. 3.** Charge density along x-axis (green line in Fig.1(a)) for bulk ZnO (a), for unrelaxed nanowire (b) and for relaxed nanowire (c). Orange rectangular shows the same area as in Fig. 1(b).

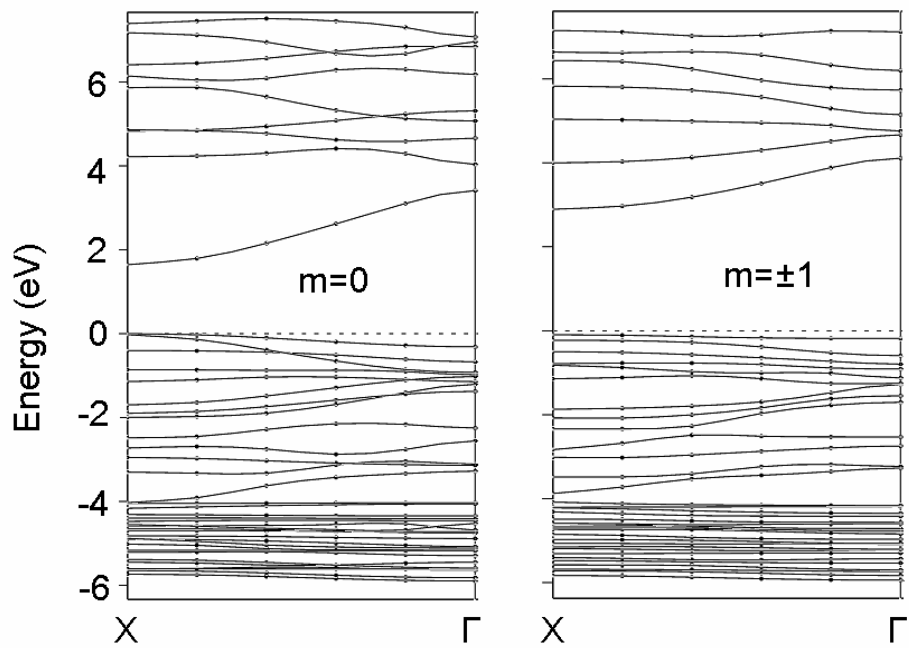

**FIG 4.** Band structure of the relaxed ZnO nanowire. Energy measured from the Fermi Surface.

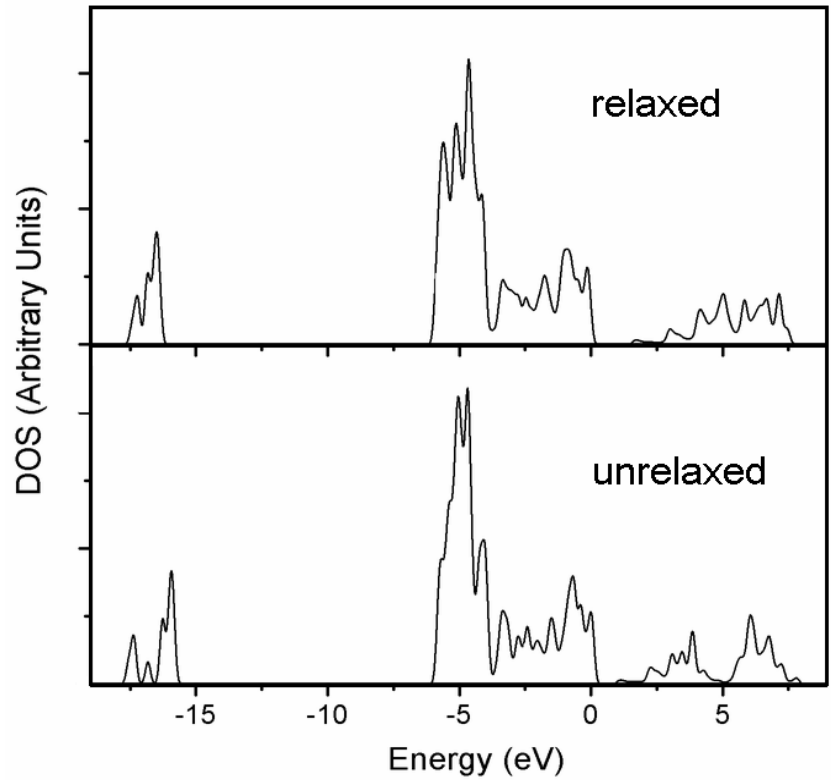

**FIG. 5** Upper panel: total DOS of the relaxed ZnO nanowire. Lower panel: total DOS of the unrelaxed ZnO nanowire. Energy measured from the Fermi surface.

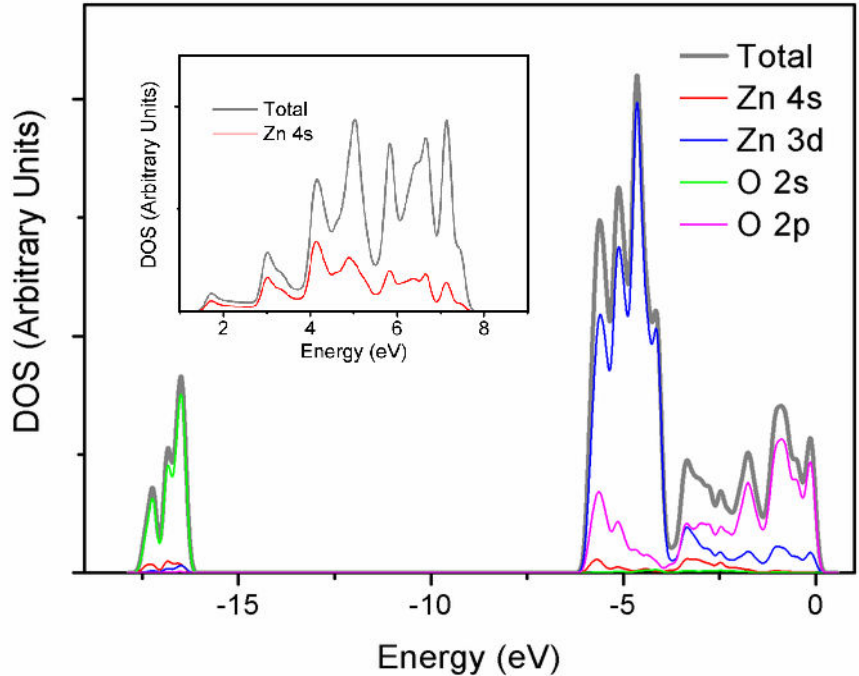

**FIG. 6.** Total and partial DOS of the valence band of the relaxed ZnO nanowire. Insert: Total and partial DOS of conduction band, only DOS of Zn 4*s* states is shown. The HO states are dominated by O 2*p* states while the LU states are mainly Zn 4*s* state.

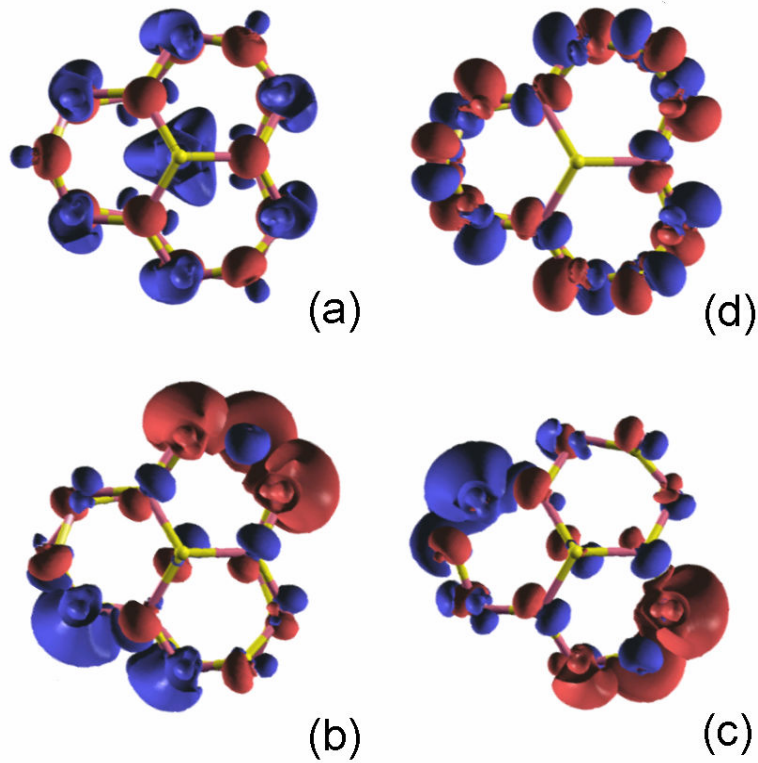

**FIG.7.** (a) Wave function of LU state, mostly made of Zn 4*s* orbits. (b)(c) Wave function of two degenerate second lowest unoccupied states, mostly made of Zn 4*s* orbits (d) Wave function of HO state, mostly from O 2*p* orbits.

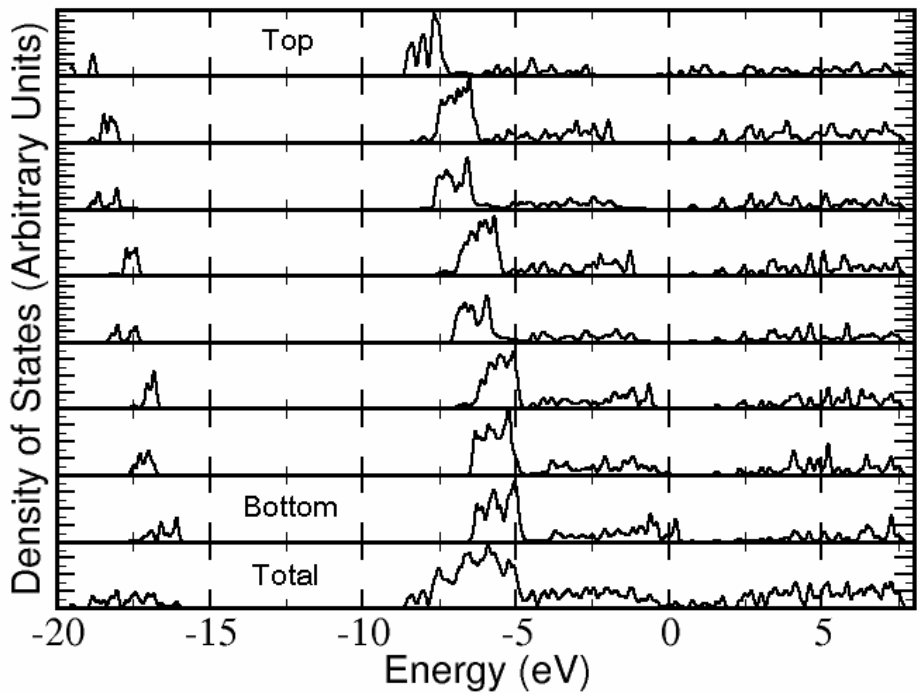

**Fig 8.** Total DOS of truncated ZnO nanowire and projections onto each layer.

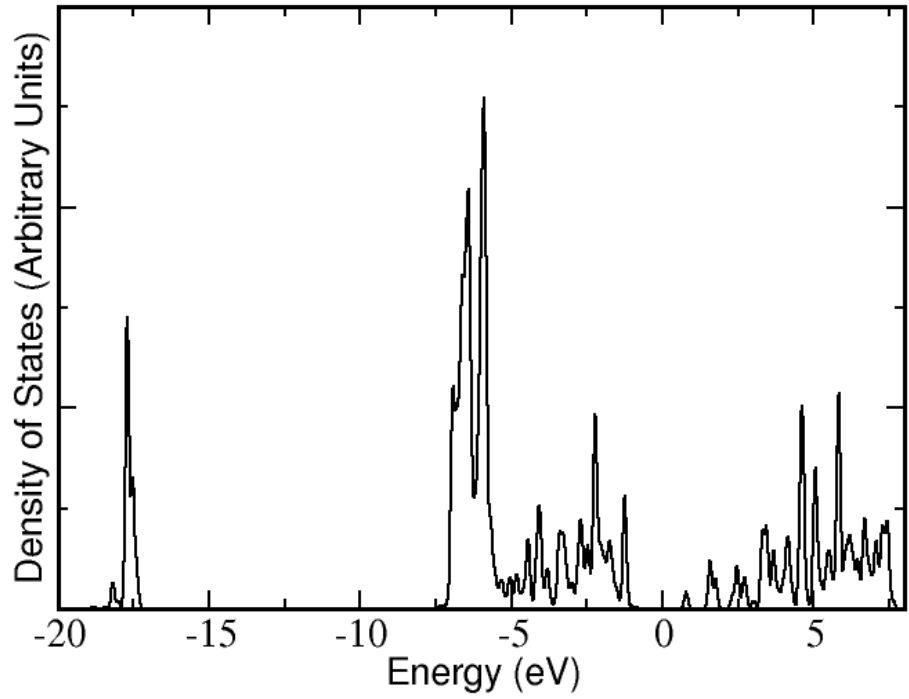

**FIG. 9.** DOS of the interior of the truncated ZnO nanowire, obtained by summing up the projected DOS of two innermost layers(panel 4 and panel 5 in Fig. 8).